\documentclass[11pt,twoside]{article}
\usepackage{newpasp}

\index{Linder. S. M.}

\begin{document}

\title{The Evolution of Ly$\alpha$ Absorbing Galaxies}
\author{Suzanne M. Linder$^1$}
\altaffiltext1{Dept.~of Physics
and Astronomy, Cardiff University, 5, The Parade, Cardiff CF24 3YB Wales, UK}
\affil{Instituto Nacional de Astrof\'isica \'Optica y Electr\'onica, Apartado
Postal 51 y 216, Puebla 72000, Pue. Mexico}

\begin{abstract}
The evolution of Ly$\alpha$ absorber counts is simulated for a
model population of absorbing galaxies.
  The distribution of gas
relative to galaxies
could evolve between moderate and low redshifts, but constraints are 
needed
on the strength and evolution of the ionizing UV
background.
\end{abstract}

Ly$\alpha$ absorber observations at low to moderate redshifts
constrain the evolution in the distribution of gas relative to galaxies. 
Absorber counts generally decrease with decreasing redshifts, although
the decrease is less rapid at redshifts less than $\sim 1.5$
(Weymann et al.~1998).  Such evolution has been explained by Dav\'e et 
al.~(1999) as largely the result of the decreasing UV ionizing background,
assuming an ionization history based upon spectra from Haardt \& Madau (1996).
The evolution in Ly$\alpha$ absorber counts ($>10^{14.3}$ cm$^{-2}$) is
shown in Fig. 1, where I assume that such absorbers arise in galaxies at 
redshift $z=0$
as simulated in Linder (2000) and Linder (1998). It is also assumed
that at at each simulated redshift a population of
galaxies with the same gaseous properties exists, and that the ionizing 
background radiation evolves as in Dav\'e et al.~(1999).  At higher redshifts
weaker absorbers are included, assuming that absorbers at a given column
density correspond to a smaller overdensity at a higher redshift, as shown
in Fig.~10 of Dav\'e et al.  Reproducing the observed evolution at high 
redshifts will require understanding both the cosmology and the large scale
process of formation of gas into galaxies.  Yet it is interesting to look
at the evolution of absorber counts due to galaxies at the lowest 
redshifts.  While qualitatively similar evolution (steeper at higher redshifts)
 is seen here as in 
Weymann et al.~(1998), one curious feature in this model
is that the absorber counts
increase from $z=0.5$ to $z=0$.

What could be happening in this redshift range?  In the simulations 
illustrated in Fig.~1 it is assumed that the total (neutral plus ionized)
gaseous extent of galaxies remains constant with redshift.  There could
be evolution as a result of the formation process of gas 
into galaxies, although it seems surprising that so much formation would
be happening at $z<0.5$.  
It is also possible that
there are as many nongalactic absorbers at $z\sim 0.5$ as at $z\sim 0$, 
where 
a given absorber is more weakly associated with a galaxy
yet more highly ionized.  In this case no evolution might be detected in 
the gaseous extent of galaxies, as for example Chen et 
al.~(2000).  
Another possibility is that the UV ionizing background could
decrease less quickly than estimated from the Haardt \& Madau (1996) models.
This seems plausible as only quasars are included in their spectra, while some 
evidence is seen that gas surrounding galaxies is ionized by the galaxies
themselves (Bland-Hawthorn et al.~1997).  Yet it is likely that there is
some evolution in the UV ionizing background and in the observable properties
of galaxies, so that a more complex evolutionary process is happening than 
that reported by Chen et al.~(2000).  One possibility 
is that absorber counts actually do increase at $z\sim 0$, where it
is most difficult to get an adequate sample of absorption line data.  In this 
case there would be more HI in the
local universe than what we have been extrapolating from absorber counts at
slightly higher redshifts.

\begin{figure}
\plotfiddle{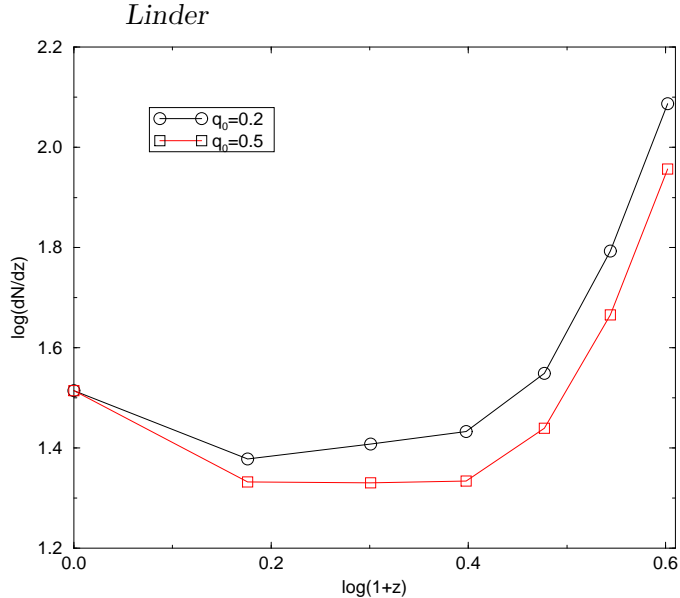}{2.6in}{270}{45}{45}{-185}{255}
\caption{The number of Ly$\alpha$ absorbers ($>10^{14.3}$ cm$^{-2}$) 
per unit redshift is shown versus redshift.  All absorbers at $z\sim 0$
 are assumed
to arise in gas extending from galaxy disks.  Absorber counts may actually
increase after $z\sim 0.5$ unless the ionizing background decreases more slowly
or the average gaseous extent of galaxies decreases rapidly.
\label{fig1}}
\end{figure}

The nature of absorbers could change quickly between moderate and low
redshifts, but at this time we know little about the strength or evolution
of the ionizing UV background.  
The strength of this background typically changes rapidly in simulations
over the range of  $z\sim 0$ to $0.8$, where observers tend to
look for the averaged relationship between galaxies and `low redshift' absorbers.

\end{document}